\documentclass[aps,pre,preprints,groupedaddress,showpacs]{revtex4}
\usepackage{graphicx}
\newcommand{\beq}{\begin{equation}}
\newcommand{\eeq}{\end{equation}}
\newcommand{\beqn}{\begin{eqnarray}}
\newcommand{\eeqn}{\end{eqnarray}}

\begin{document}


\title{Mechanical properties of the two-filament insulin amyloid fibril: a theoretical study}
\author{Chiu Fan \surname{Lee}}
\email{C.Lee1@physics.ox.ac.uk}
\affiliation{Physics Department, Clarendon Laboratory,
Oxford University, Oxford OX1 3PU UK}
\date{\today}

\begin{abstract}
We study the two-filament insulin fibril's structure by incorporating recent
simulation results and mechanical measurements. 
Our investigation suggests that the persistence length measurement correlates well with the previously proposed structural model, while the  elasticity measurement suggests that stretching the fibril may involve hydrogen bond breakage. 
Our work illustrates an attempt to correlate nanoscale measurements with
microscopic information on the quaternary protein structure.
\end{abstract}


\maketitle


\section{Introduction}
Amyloids are insoluble fibrous protein aggregations stabilized by a network 
of hydrogen bonds and hydrophobic interactions \cite{Sunde97, Dobson03, Radford00,Sawaya07}. They are 
intimately related to many neurodegenerative diseases such as the 
Alzheimer's Disease, the Parkinson Disease and other prion diseases.
Better characterization of the various properties of amyloid fibrils is 
therefore of high importance for further understanding their associated 
pathogenesis. Here
we study the Two-Filament Insulin Fibril (TFIF) by incorporating 
the simulation results in \cite{Park06}, the gaussian network model for amyloid fibrils introduced in \cite{Knowles07b}, and the measurements on the mechanical properties performed in \cite{Smith06}. Specifically, we consider the implications of the elasticity and persistence length measurements performed in \cite{Smith06} on the TFIF's structure.

\begin{figure}
\caption{A schematic of the helical insulin fibril investigated in 
\cite{Jimenez02}. The fibril consists of two 
filaments (red and blue) with a cross-beta sheet core structures (the light blue lines) and an amorphous layer surrounding them (the grey 
areas).}
\label{Fibril}
\begin{center}
\includegraphics[scale=.4]{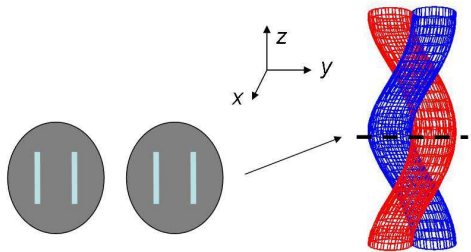}
\end{center}
\end{figure}

An amyloid fiber consists of intertwining filaments with a cross-beta sheet 
core structure \cite{Dobson03}. Within the cross-beta sheet core, the beta 
strands are separated by 4.8{\AA} along the fibril axis. With the help of 
the carefully determined electrons density map for insulin fibrils in 
\cite{Jimenez02}, we have a relatively good idea on the location of the 
cross-beta sheets in the TFIF (c.f. 
Fig.~\ref{Fibril}). Specifically, experimental observations \cite{Bouchard00,Jimenez02} 
suggest that (i) the beta sheets are in the parallel 
configuration; (ii) 
there is a gap of 10{\AA} between the two layers of beta-sheets within each
filament; and (iii) the electrons density map suggests that there is a gap of about 30{\AA} between the two filaments.
These facts suggests that a structural model in which
the TFIF is a doube-helix where each filament 
consists of two layers of cross-beta sheets (c.f. Figs.~1 and 
2).

\section{Persistence length}
In this section, we discuss the persistence length of the TFIF.
The formalism developed in \cite{Panyukov00,Panyukov00b} indicates that for a `rod-like' helical ribbons, the persistence length, $L_p$, is of the form:
\beq
L_p =  \frac{1}{k_BT}\left[\frac{1}{2K_B}+ \frac{1}{2K_S} \right]^{-1}\ ,
\eeq
where $K_B$ is the bending stiffness and $K_S$ is the splaying stiffness of the ribbons. For the cross-section indicated by the black line in Fig. 1,  $K_B$ refers to
bending about the $x$-axis and $K_S$ refers to bending about the $y$-axis. 
Due to the gap of $\sim  30${\AA} between the two filaments in the $y$ direction (c.f. Fig.~2), we expect that bending about the $x$-axis is much 
more difficult than bending about the $y$-axis, i.e., $K_B \gg K_S$. 
 We will therefore further simplify the expression for $L_p$ to $2K_S/k_BT$. 

To get an idea on the magnitude of $K_S$ for a generic two-layer beta-sheet, we look at the simulation results in \cite{Park06} on a bilayered anti-parallel beta-sheets formed with a 16-amino-acid peptides (RAD16). In \cite{Park06}, the splaying stiffness for RAD16 is found to be about $(1.47 -2.22) \times 10^{-25}$Nm$^2$ \cite{Park06}. As we have stipulated that the splaying stiffness dictates the persistence length,  the persistence length of $22\pm 3 \mu$m as measured in \cite{Smith06} suggests that the splaying stiffness for TFIF is about $4.6 \times 10^{-26}$Nm$^2$. In other words, the $K_S$ for TFIF is approximately $(21-30)\%$ of that of RAD16, which in turn suggests that the width of the TFIF's beta-sheet is about $(47\% -54\%)$ of that of RAD16, i.e., the TFIF's beta-sheet is 7.6 to 8.6 amino-acides in length. This correlates reasonably well with the structure promoted in \cite{Jimenez02} where the beta-sheet core structure consists of alternating insulin A chain (21 amino-acid long) and B chain (30 amino acid long). The above comparison is of course only valid if the hydrogen bond networks are solely responsible for the elastic properties of TFIF. This assumption is supported by simulation studies in \cite{Park06} and experiments \cite{Knowles07b}. Another caveat here is that the RAD16 has an anti-parallel beta-sheet structure while the TFIF has a parallel beta-sheet structure. We have therefore implicitly assumed that the elastic properties is independent of the beta-sheet directional arrangement. This assumption is again supported by previous theoretical work \cite{Chou85} and recent experimental work \cite{Knowles07b}.

\begin{figure}
\caption{A schematic of the beta sheet arrangement at the cross section 
indicated by the black line in Fig.~\ref{Fibril} of the helical fibril. The 
green panels depict the beta sheets and the red lines represent the 
hydrogen bonds stabilizing the beta sheets. All numerical values are in 
units of {\AA}.}
\label{Strands}
\begin{center}
\includegraphics[scale=.4]{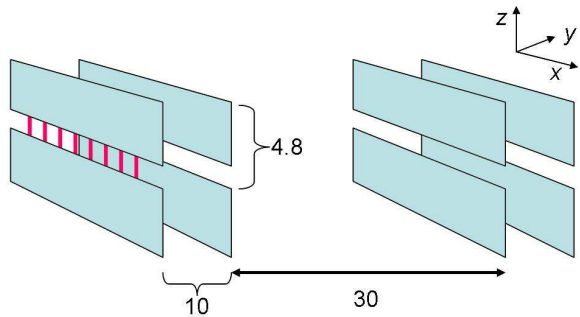}
\end{center}
\end{figure}

\section{Elasticity}
In \cite{Knowles07b}, a Gaussian network model (GNM) is introduced to account for the general mechanical properties of amyloid fibril. In the GNM, the spring constant of each hydrogen bond, $k_H$, is taken to be about 12N/m \cite{Chou85}.  This spring constant estimate is corroborated by the simulation results in \cite{Park06} where the stretching stiffness is found to be about $K_T \simeq (1.19 - 1.51) \times 10^{-7}$N for RAD16, which  suggests that each hydrogen bond has a spring constant of $7.7-9.8$N/m. For the TFIF, if we assume that i) the spring constant for an hydrogen bonds is 10N/m, ii) there are 8 hydrogen bonds per cross-beta sheet (c.f. the previous section), iii) the area of cross section is $14$nm$^2$ \cite{Smith06}, and iv) the separation between each array of intersheet hydrogen bonds is 4.8{\AA}. With these assumptions, the Young's modulus of the TFIF can be estimated to be about 13GPa. This is close to the value $17.2 \pm 1.5$GPa, which is suggested to be the Young's modulus for a generic amyloid fibril based on the GNM \cite{Knowles07b}. On the other hand, the measured value for the TFIF's  Young's modulus by atomic force microscope experiment is $3.3\pm 0.4$GPa \cite{Smith06}. Besides the difference in the predicted and measured Young's modulus, 
it is found in the simulation studies in \cite{Park06} that the stretching elasticity of the RAD16 amyloid fibril  is only linear within a very short range. Specifically, the extension elasticity is only linear when the extension is less than 0.1\% of the original length. On the other hand, the 
AFM experiments performed in \cite{Smith06} suggests that the elastic measurements are performed when the fibril is extended by more than 5\% of its original length. These discrepancies may be explained by hydrogen bonds breakage under extension. To validate this picture and to further elucidate the exact nature of amyloid fibril's elasticity, further experimental and theoretical work is required. 

\section{Conclusion}
We have investigated the TFIF's structure by incorporating 
the simulation results in \cite{Park06}, the GNM for amyloid fibrils introduced in \cite{Knowles07b}, and the measurements on the mechanical properties performed in \cite{Smith06}. Our comparison studies suggest that the persistence length measurement correlates well with the structural model introduced in \cite{Jimenez02}, while the stretching elasticity measurement suggests that the extension of the fibril may involve complicated mechanism such as hydrogen bond breakage.

\begin{acknowledgements}
The author thanks the Glasstone Trust and Jesus College (Oxford) for 
financial support.   
\end{acknowledgements}

\bibliography{Partition}
\end{document}